\documentclass[preprint,12pt]{elsarticle}

\usepackage[utf8]{inputenc}
\usepackage{amsmath,amssymb,amsfonts}
\usepackage{graphicx}
\usepackage[mathscr]{eucal}
\usepackage{hyperref}
\usepackage{times}
\usepackage{comment}

\journal{Mathematical Biosciences}

\begin{document}

\begin{frontmatter}
\title{A Topological Data Analysis Study on Murine Pulmonary Arterial Trees with Pulmonary Hypertension}

\author[1,2]{Megan Chambers}
\author[1,3]{Natalie Johnston}
\author[1]{Ian Livengood}
\author[1]{Miya Spinelli}
\author[1]{Radmila Sazdanovic}
\author[1]{Mette S Olufsen\corref{cor}}

\affiliation[1]{organization={North Carolina State University},
            addressline={2311 Stinson Drive}, 
            city={Raleigh},
            postcode={27695}, 
            state={NC},
            country={USA}}
\affiliation[2]{organization={Virginia Military Institute},
            addressline={319 Letcher Avenue}, 
            city={Lexington},
            postcode={24450}, 
            state={VA},
            country={USA}}
\affiliation[3]{organization={Duke University},
            addressline={415 Chapel Drive}, 
            city={Durham},
            postcode={27708}, 
            state={NC},
            country={USA}}
            
\begin{abstract}
Pulmonary hypertension (PH), defined by a mean pulmonary arterial blood pressure above 20 mmHg, is a cardiovascular disease impacting the pulmonary vasculature. PH is accompanied by vascular remodeling, wherein vessels become stiffer, large vessels dilate, and smaller vessels constrict. Some types of PH, including hypoxia-induced PH (HPH), lead to microvascular rarefaction. The goal of this study is to analyze the change in pulmonary arterial network morphometry in the presence of HPH using novel methods from topological data analysis (TDA). In particular, we employ persistent homology to quantify arterial network morphometry for control and hypertensive mice by characterizing arterial trees extracted from micro-computed tomography (micro-CT) images.

To compare results between control and hypertensive animals, we normalize generated networks using three pruning algorithms. This proof-of-concept study shows that the pruning method affect the spatial tree statistics and complexity of the trees. Results show that HPH trees have higher depth and that the directional complexities correlate with branch number, except for trees pruned by vessel radius, where the left and anterior complexity are lower compared to control trees. While more data is required to make a conclusion about the overall effect of HPH on network topology, this study provides a framework for analyzing the topology of biological networks and is a step towards the extraction of relevant information for diagnosing and detecting HPH.
\end{abstract}


\end{frontmatter}

\section{Introduction}

\noindent Cardiovascular diseases are the leading cause of death in the western world. The World Health Organization (WHO) \cite{WHO21} estimates that in 2019 17.9 million people died from cardiovascular diseases, approximately 32\% of all deaths. While cardiovascular diseases encompass many pheno-types, a common trait is that they are associated with remodeling of the vasculature and the heart, causing a range of functional problems, the most prominent being high blood pressure. Another important finding is that early detection is essential to improve quality of life and lessen the burden on the healthcare system \cite{WHO21}. Cardiovascular diseases are typically diagnosed by combining imaging measurements, such as Computed Tomography (CT) and Magnetic Resonance Imaging (MRI) with dynamic, e.g.,  heart rate, blood pressure and flow, and static measurements. The latter including patient characteristics as well as physiological and biochemical markers dervied from blood tests. Cardiovascular disease can target any part of the vascular system changing blood pressure, blood flow, and/or heart rate often caused by tissue remodeling at the local (within specific vessels) and global (network) levels. Tissue remodeling has two effects: changes in compliance within vessels and changes in network morphometry, both impact blood pressure and flow. This study uses techniques from topological data analysis to characterize changes in the morphometry of pulmonary arteries in mice with hypoxia-induced pulmonary hypertension. 

Pulmonary hypertension (PH) is characterized by high blood pressure (a mean $\geq$ 20 mmHg) in the main pulmonary artery. The severe increase in pulmonary arterial pressure is typically caused by vascular remodeling and inflammation in the veins or left heart \cite{Beshay21}. Symptoms of the disease include shortness of breath, fatigue, dizziness, chest pain, heart palpitations, and swelling of the legs and ankles. These symptoms are common in many illnesses, making PH difficult to diagnose \cite{Galie15}. Moreover, PH is a heterogeneous disease encompassing five subtypes, each with a separate pathophysiology \cite{Simonneau19}. However, early diagnosis and targeted treatment can improve quality of life by delaying severe complications, which is essential because all but one type of PH has no cure \cite{Lau15}. This study analyzes the morphometry in murine pulmonary arterial networks excised from healthy and hypoxia-induced pulmonary hypertension animals. In PH induced by hypoxia (HPH) vascular remodeling impacts the pulmonary arteries \cite{Molthen04,Vanderpool11}. The disease starts in the arterioles, stiffening and constricting the vessels, then migrating to the larger arteries, which stiffen and dilate. Hopkins et al. \cite{Hopkins02} have characterized the structural changes in individual vessels, but less is known about how the network morphometry changes, the main focus of this study,  changes.

\medskip
\noindent \textbf{Pulmonary arterial morphometry} characterizes the network properties in space. Since early contributions by Murray in 1926 \cite{Murray26}, many researchers (e.g., \cite{Chambers20,Horsfield77,Molthen04,Singhal73,Vanderpool11}) have examined the morphometry of the pulmonary arterial network. The early studies by Murray \cite{Murray26}, and Zamir \cite{Zamir78} devise optimality principles, describing arterial branching that minimize pumping power and lumen volume. These studies assume that the arterial network bifurcates and that the dimensions of the two daughter vessels can be determined as functions of the parent vessel. Results from these studies include a power law defining how vessel radii change across a bifurcation, an asymmetry ratio relating the radii of the two daughter vessels, and an area ratio relating the combined cross-sectional area of the daughter vessels to that of the parent vessel. These studies were supplemented by work from Singhal et al. \cite{Singhal73}, Horsfield \cite{Horsfield77}, and \cite{Olufsen99,Olufsen00} that incorporate data from lung casts to devise relations between parent and daughter vessels. The casts are generated by injecting liquid resin into the arterial networks. The vessel dimensions are measured using calipers on the hardened resin cast. The data provide geometric information, but each study only examined a single lung, as human cadaver data are not easily obtained. Moreover, the casts are fragile, and there is inherent human error in using calipers to measure the vessel dimensions. More recent studies by Molthen et al. \cite{Molthen04}, Vanderpool et al. \cite{Vanderpool11}, Davidoiu et al. \cite{Davidoiu16}, and Chambers et al. \cite{Chambers20} use medical imaging to accurately and efficiently extract geometric information from arterial networks.

\medskip
\noindent \textbf{Persistent homology} refers to a TDA technique wherein combinatorial structures are successively built from a data set, and their \emph{homology} is used to define descriptors of the shape of the data. These novel techniques have been used to analyze the arterial networks in the brain \cite{Bendich16} and bronchial networks \cite{Belchi18}, but to our knowledge, this study is the first to use persistent homology to characterize pulmonary arterial networks. Bendich et al. \cite{Bendich16} analyze human brain arterial trees generated from a tube-tracking segmentation algorithm on magnetic resonance images (MRI).  They find that the 0--dimensional persistence correlates strongly with age, and dimension one with sex.

Persistent homology is also used to  compare airway CT images from healthy and chronic obstructive pulmonary disease (COPD) patients \cite{Belchi18}. This study found that 0--dimensional persistent homology can distinguish the patient groups, while 2--dimensional persistent homology only detects inspiration and expiration. 

Motivated by these studies, 
we use persistent homology with labeled spatial trees extracted from micro-CT images. We compute the 0--dimensional persistent homology with respect to the \emph{height filtration} in $\mathbf{R}^3$ in the forward and reverse directions. As a result we get six persistence diagrams per spatial  tree  and the total persistence in each direction, known as the \emph{directional complexity}. These numerical values provide a way to compare control and HPH networks. For example, we find that tree  depth is larger in the HPH trees, directional complexity correlates with the branch count, and that the left and dorsal complexities are lower in HPH. 

\section{Methods}

\noindent Our objective is to identify 
topological and numerical summaries/features
that can characterize differences between control and HPH arterial networks extracted from micro-CT images. 
The methods include into three major steps.
First, we generate a 3-dimensional (3D) rendered network from the micro-CT image, constructing a labeled spatial tree with \emph{edges}, representing vessels, and \emph{vertices}, representing junctions. Each vertex labeled by its length, spatial orientation, and radius has coordinates in $\mathbb{R}^3$. Second, we compare the spatial trees obtained from control and HPH animals. To get biologically significant information from these spatial trees, comparing only corresponding parts of the control and HPH trees is essential. In HPH, the diameter of the large vessels increases \cite{Chambers20}, making more vessels visible in the 3D rendered network. Therefore, to compare the networks, we employ
pruning algorithms to normalize the trees.
Finally, we use persistent homology to characterize the normalized networks. Inspired by Belchi et al. \cite{Belchi18}, we -compute 0-dimensonal persistent homology of the point cloud corresponding to spatial trees with respect to the height filtration and corresponding  directional complexities to be used as numerical descriptors in comparison.

\begin{figure}[htb!]
    \centering
   \includegraphics[width=\textwidth]{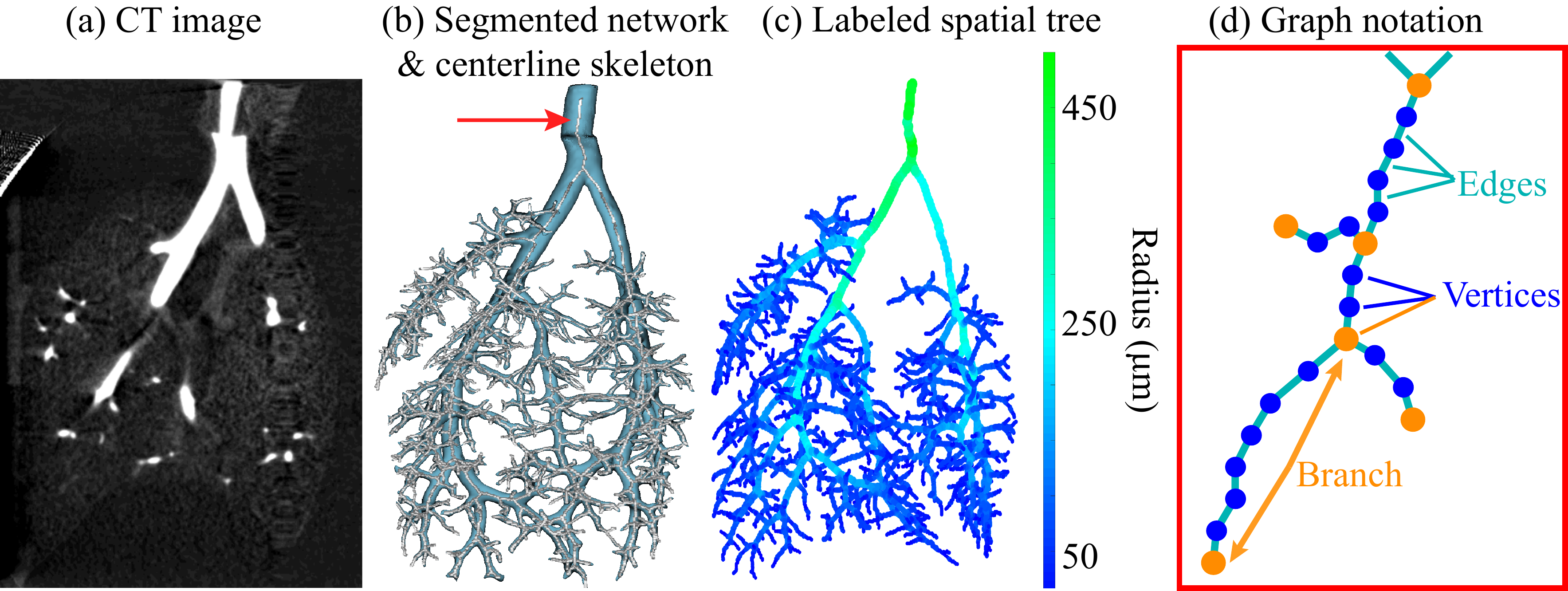}
    \caption{Spatial tree extraction process. (a) Example micro-CT image from a control mouse (coronal view). (b) 3D rendering of the segmented arterial tree overlaid with the centerline skeleton (marked with the red arrow). (c) Labeled spatial tree extracted from the skeleton, with branch colors denoting the vessel radii. (d) Detailed representation of edges and vertices in the labeled spatial tree. The edges (teal) connect two vertices (blue). Vertices forming a junction (connected to three edges) or creating a root or terminal point (connected to one edge) are marked in orange. The series of vertices and edges between 2 junction points or between a junction point and a terminal point is called a \emph{branch}.}
    \label{fig:seg}
\end{figure}

\subsection{Imaging protocol}
\begin{figure}[ht]
    \centering
    \includegraphics[width=\textwidth]{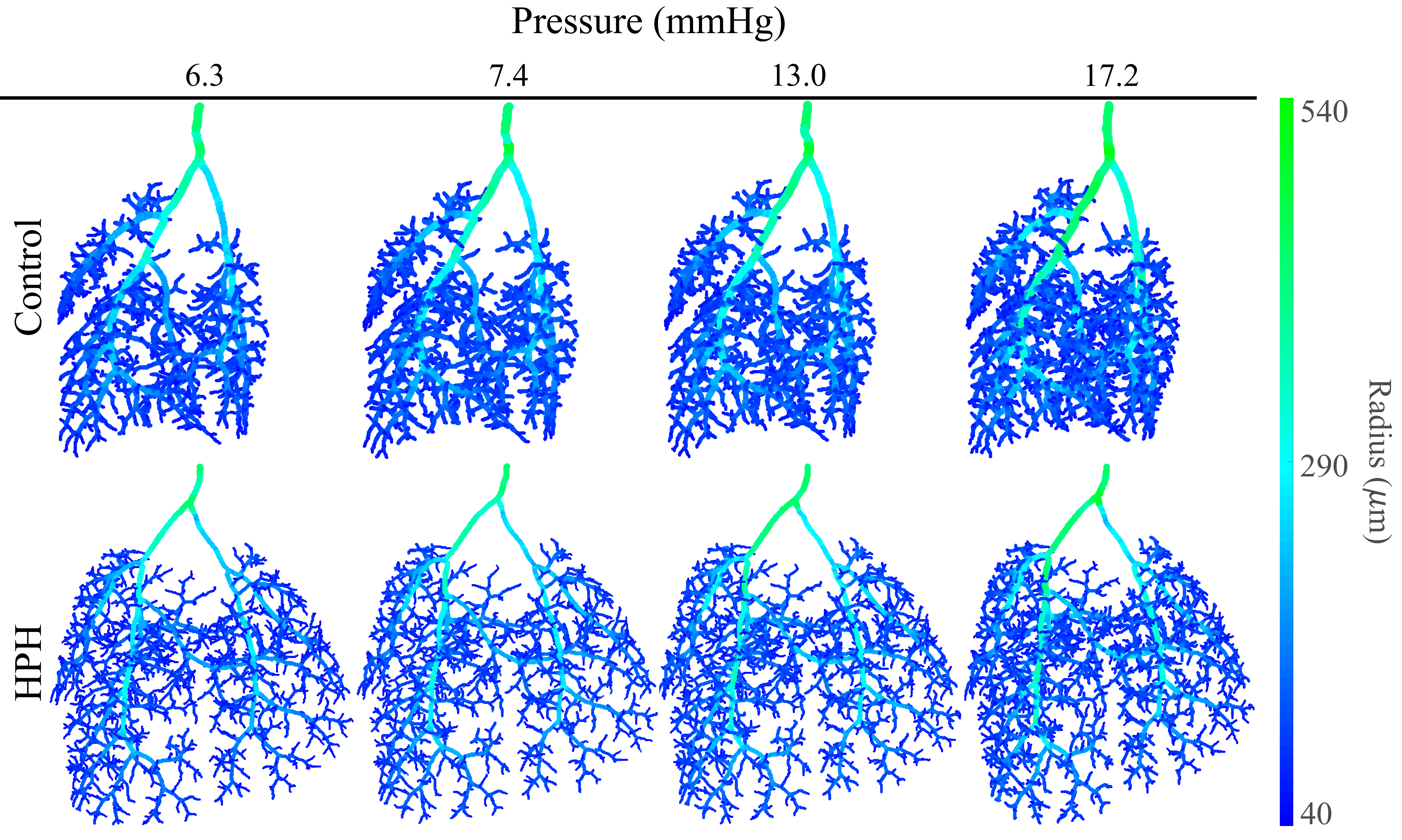}
    \caption{Labeled spatial trees from a control mouse (top row) and an HPH mouse (bottom row) extracted from lungs perfused at four different pressures in the main pulmonary artery (6.3, 7.4, 13.0, and 17.4 mmHg). As the pressure increases, the contrast is transported further, making smaller vessels visible in the image and resulting in more branches in the spatial trees.}
    \label{fig:pressures}
\end{figure}

\noindent The micro-CT images used in this study were made available by Chesler, University of California-Irvine \cite{Vanderpool11}. This study uses images from six C57BL6/J mice, age 10-12 weeks and weight 25.8 $\pm$ 2.3 grams. Of the six mice, 3 are controls and 3 have PH induced by placing them in an hypoxic environment (FiO2 reduced by half) for 10 days. The mice were mediated with (52 mg/kg body weight) pentobarbital sodium and euthanized by exsanguination before the lungs were extracted. After extraction, the lungs were imaged following the protocol described in \cite{Vanderpool11}. First, a cannula (PE-90 tubing, 1.27 mm outer and 0.86 mm inner diameter) was positioned in the main pulmonary artery (MPA) well above the first arterial bifurcation. After, the lungs are ventilated in a gas mixture with (15\% O$_2$-6\% CO$_2$, balance nitrogen), rinsed with a physiological salt solution, and perfused with perfluorooctyl bromide (PFOB).  The lungs were further prepared by periodically adjusting the intravascular pressure from 0 to 25 mmHg multiple. After preparation, the arterial pressure was kept constant as the lungs were rotated around an X-ray beam at $1^\circ$ increments to obtain 360 planar images. The lungs were imaged at main pulmonary arterial pressures of 6.3, 7.4, 13.0, and 17.4 mmHg (Figure \ref{fig:pressures}). The planar images for each pressure were  reconstructed using the Feldkamp cone-beam algorithm and converted into a 3D volumetric dataset, saved as a Digital Imaging and Communications in Medicine (DICOM) 3.0 image (Figure \ref{fig:seg}(a)). 

\subsection{Segmentation and skeletonization}\label{sec:segmentation}

\noindent The micro-CT images are saved in the (DICOM) 3.0 format. Each image has the dimensions $\left(497\times497\times497\right)$ voxels with a spatial resolution of 30-40 $\mu$m per voxel. Each voxel $v$ is represented by a voxel complex with spatial coordinates $(x_v, y_v, z_v)$ and intensity $I(v) \in [0, 255]$ with 0 representing the intensity of a black voxel and 255 of a white voxel.

\medskip
\noindent\textbf{Segmentation} is the process used to partition the voxels into the set of \emph{foreground} voxels, which represent the arteries, and \emph{background} voxels representing the surrounding tissue. We segment the images using the open-source program 3D Slicer \cite{Fedorov12, Kikinis14, 3DSlicer}, Kitware, Inc., employing a combination of segmentation techniques, including global thresholding, median smoothing, and manual editing as described in detail in our previous study \cite{Chambers20}. Global thresholding is used to identify voxels in the foreground, finding voxels with intensities $I(v)\in [\tau_{\text{min}}, \tau_{\text{max}}]$. The pulmonary arteries are the only anatomical structures in the image, so $\tau_{\text{max}} = 255$ for all images. The minimum threshold $\tau_{\text{min}}$ is adjusted ad hoc to ensure that all visible arteries are included.  To reduce noise, median smoothing is used, replacing the intensity of all the voxels within a kernel of $(3\times3\times3)$ voxels with the median of the adjacent voxel intensities. Since the cannula and the hypobaric cylinder can distort the image, manual editing is required to remove voxels from the foreground, which do not represent the true arterial structure. The result of segmentation is a foreground which can be visualized as a 3D rendered surface representing the pulmonary arterial tree. The constructed foreground is referred to as the \emph{segmented network} (shown in Figure~\ref{fig:seg}(b)). Each \emph{segmented artery} is a portion of the segmented network that lies between two junctions.

\medskip
\noindent \textbf{Skeletonization:} The \emph{skeleton} is a voxel complex comprising a thinned representation of the segmented arteries. In the skeleton, the branching structure of the segmented arteries is preserved, i.e., each branch is represented by a set of voxels that are one voxel in width and centered in the artery. The skeleton is obtained by iteratively removing voxels from the segmented arteries using Couprie and Bertrand’s ``Asymmetric Thinning" algorithm. For details see \cite{Chambers20, Couprie16}. Example centered skeleton is shown in Figure \ref{fig:seg}(b).

\medskip
\noindent \textbf{Distance map:}  To obtain the dimensions for the segmented arteries, we generate a \emph{distance map}, an associated voxel complex with the same spatial dimensions as the original image. For each voxel $v$, we compute a distance voxel $v_D$ denoting the distance from $v$ to the nearest background voxel $u$ computed as 
\begin{equation} 
v_D=\min_{u\in \text{background}}\left\Vert u-v \right\Vert_2,
\label{eq:dmap9}
\end{equation}
where  $\left\Vert u-v \right\Vert_2$ is the Euclidean distance from $v$ to each background voxel, $u$, and the distance voxel $v_D$ has the same spatial coordinates $(x_v,y_v,z_v)$ and intensity $I(v_D)=d(v)$ as the voxel $v$. This distance map, the skeleton, and the labeled spatial tree are obtained using the Spatial Graph Extractor (SGEXT) in the Digital Geometry Tools and Algorithms Library (DGtal) available in GitHub \cite{DGtal18,Sgext,Hernandez18}.
 
\subsection{Spatial tree labels}

\noindent Each skeleton is used to generate a \emph{spatial graph} (also known as an embedded metric graph), a collection of vertices in $\mathbb{R}^3$ connected by edges. Every voxel in the skeleton corresponds to a vertex in the spatial graph. Each vertex, denoted by a numerical ID $A$, has 3D coordinates $(x_A,y_A,z_A)$. Each edge connected by 2 vertices $A$ and $B$ is denoted $e_{AB}$. If $\deg(A)=1$, $A$ is either at the inlet to the MPA (called $A_\text{root}$) or a \emph{leaf}, a terminal vertex. If $\deg(A)=2$, $A$ is a vertex along the length of a segmented artery. If $\deg(A)>2$, $A$ is a \emph{junction point}, a point where a \emph{parent} artery splits into multiple \emph{daughter} arteries. Typically, junction points have degree 3, as 98-99\% of junctions are bifurcations \cite{Chambers20}. All edges are oriented toward blood flow (away from $A_\text{root}$). With this orientation in place, we can refer to $e_{AB}$ as an edge from start vertex $A$ to end vertex $B$. 

Constructing the spatial graph is subject to errors that can be corrected using the protocol introduced in~\cite{Chambers20}. Errors include edges that do not accurately represent the segmented arteries, small cycles that arise when 3 adjacent voxels are connected in a loop, duplicated edge points, and duplicated edges that connect the same vertices. False branches must be manually identified and removed from the graph, while small cycles and duplicate edges/points can be removed automatically. Cycles are broken by removing the longest edge in the cycle. After implementing the corrections, the graph becomes a \emph{spatial tree} $T$ (Figure \ref{fig:seg}(c)).

Each vertex in $T$ is labeled with its radius, which is determined from the distance map. Recall that the skeleton is centered in the segmented arteries. Therefore, for voxel $v$ in the skeleton, the measurement $D_v$ calculated using equation (\ref{eq:dmap9}) gives an estimate for the radius of the artery centered at $(x_v, y_v, z_v)$. 
Hence, every vertex $A$ in $T$ can be labeled by its radius $r_A=d(v)$, i.e., for the corresponding voxel $v$ $(x_v, y_v, z_v)=(x_A, y_A, z_A)$.

In the 3D rendered surface, every artery is represented by a collection of edges connecting consecutive degree 2 vertices between either 2 junctions or a junction and a leaf. We refer to the collection of edges as a branch of $T$, i.e., each branch corresponds to an artery. Branches are labeled with an average radius and length in microns ($\mu$m). To convert the voxel measurements into $\mu$m, all dimensions, $x$, $y$, and $z$, are multiplied by the scaling factor $\lambda_T$, obtained by dividing the cannula's radius (430 $\mu$m) by the average of the first 5 radius measurements of the MPA, the radii of vertices $A_{\text{root}}, A_1, ..., A_4$, i.e.,
\begin{equation}
    \lambda_T = \frac{5\times430\text{ } \mu \text{m}}{(r_{A_{root}}+r_{A_1}+r_{A_2}+r_{A_3}+r_{A_4}) \text{ voxels}}.
    \label{eq:scale}
\end{equation}
The radius $r_{AB}$ of a branch from a leaf/junction point $A$ to a leaf/junction point $B$ is represented by the interquartile mean (IQM) of the radius measurements from $A$ to $B$, and the degree 2 vertices $A_1, ..., A_m$ along the branch,a
\begin{equation} 
r_{AB}=\lambda_T  \hspace{1mm}(IQM\left(r_A, r_{A_1}, r_{A_2} ..., r_{A_m}, r_B\right) ).
\label{eq:radius}
\end{equation}
\noindent The length $L_{AB}$ of a branch is  the scaled sum of the Euclidean distances between consecutive vertices along that branch,
\begin{equation} 
L_{AB}= \lambda_T\left(\left\Vert A-A_1 \right\Vert_2 +  \sum_{i=2}^{m} \left\Vert A_{i-1}-A_i \right\Vert_2 + \left\Vert A_m-B \right\Vert_2\right).
\label{eq:length}
\end{equation}

\subsection{Spatial tree statistics}

\begin{figure}[b!]
    \centering
    \includegraphics[width=\textwidth]{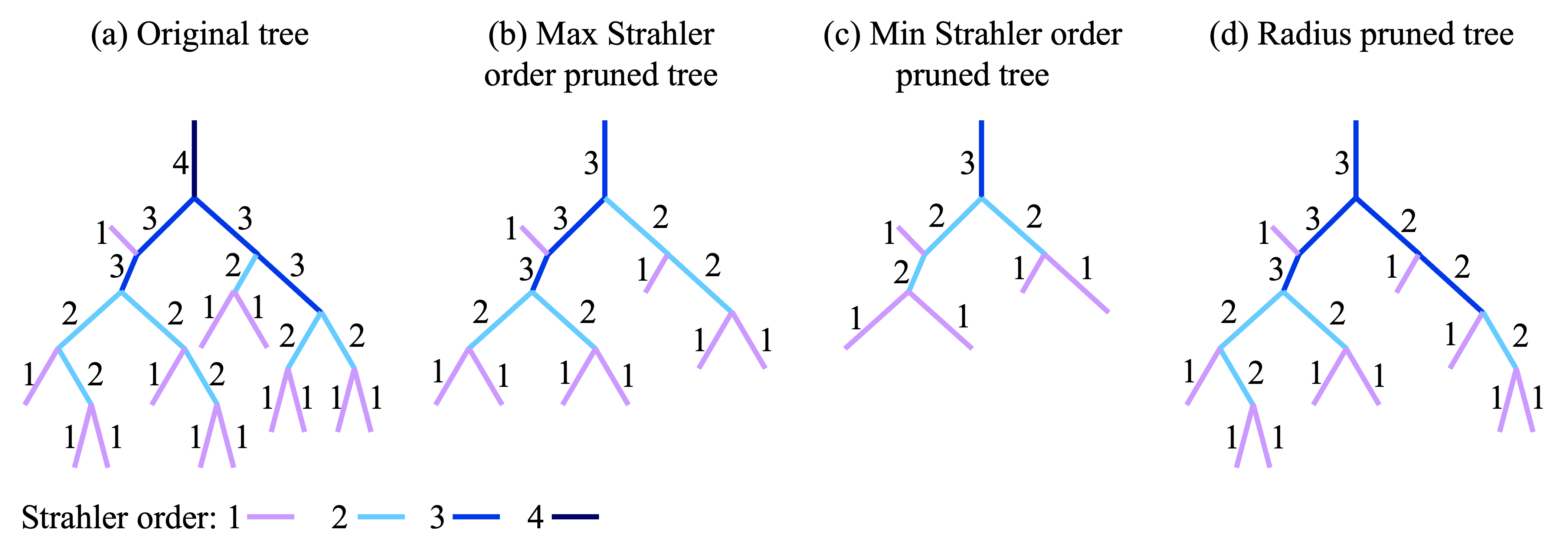}
    \caption{The Strahler ordering system illustrated on an example tree. (a) Example tree of Strahler order 4 to be pruned. (b) Max Strahler order pruning to order 3. This tree removes as few vessels as possible from the original tree to reach Strahler order 3. (c) Min Strahler order pruning to order 3. This tree removes as many vessels as possible still getting a tree of Strahler order 3, i.e., if one more round of pruning was performed the tree will have Strahler order 2. (d) Radius pruned tree. This tree removes vessels with radius smaller than a given threshold. Note the pruned tree is also Strahler order 3, but has different structure than any of the Strahler order pruned trees.}
    \label{fig:Strahlerorder}
\end{figure}

\noindent
The number of branches, leaves, tree depth and the Strahler order are used as features of spatial trees. 
The \emph{tree depth} is defined  as the number of branches in the longest direct path from $A_\text{root}$ to any leaf. The \emph{Strahler order} ($SO$) of a branch is an indicator of its level within the tree \cite{Vanbavel92}. Terminal branches all have $SO=1$. If two joining daughter vessels have equal $SO$ (i.e., $SO_{d1} = SO_{d2}$), then the $SO$ parent edge is equal to $SO_p = SO_{d1} + 1$.  Otherwise, the $SO$ of the parent edge is $SO_p = \max{\{SO_{d1}, SO_{d2}\}}$ (Figure \ref{fig:Strahlerorder}).  The Strahler order of a tree refers to the maximum Strahler order of any branch within that tree. For the trees examined here, this will always be the root branch, the MPA, denoted $SO_{MPA}$. 

\subsection{Spatial tree pruning} \label{sec:prune}
      
\noindent 
To distinguish HPH mice from controls  we use pruning techniques to obtained comparable trees. This was necessary because the trees for HPH mice had too many branches compared to the control since the corresponding vessels are wider and therefore detectable in the micro-CT images. If pulmonary trees were symmetric tree depth would be sufficient for their characterization and pruning would be straightforward, but pulmonary arterial networks are asymmetrical \cite{Chambers20,Murray26,Olufsen00,Zamir78}.
 In this study, we propose 3 pruning methods wherein terminal branches are systematically removed. Two pruning techniques rely on computing the Strahler order, and the third method uses the vessel radius.

\medskip
\noindent \textbf{Maximum Strahler order pruning} starts at the leaves removing sister branches at the lowest Strahler order, branches for which both daughter vessels in a junction are of Strahler order 1 (i.e., $SO_{d1} = SO_{d2}=1$). Once all branches of this type has been removed, the Strahler order of the tree is reduced by one. Pruning continues until the desired Strahler order is obtained. This pruning method is illustrated in Figure~\ref{fig:Strahlerorder}. Panel (a) shows the original tree and (b) the maximum Strahler order pruning, removing all sister vessels with Strahler order 1 from the original network (a). The new Maximum Strahler order pruned tree has exactly one Strahler order lower than the original tree. 

\medskip
\noindent \textbf{Minimum Strahler order pruning:} Trees at any given Strahler vary significantly in the number of branches and tree depth. To eliminate this variation we also perform minimum Strahler order pruning. This pruning method
removes all pairs of terminating sister branches for which $SO_{d1} = SO_{d2} = 1$. In addition, this algorithm removes any branches that will not impact the final Strahler order.  Strahler orders are recomputed, and the process continues until the next iteration of branch removal gives the minimal tree with the desired Strahler order. This method is illustrated in Figure \ref{fig:Strahlerorder}(c).

\medskip
\noindent \textbf{Radius pruning: }
This pruning technique involves removing pairs of terminal daughter branches with a radius less than or equal to a given radius threshold $\tau_r$. This threshold is  chosen such that any group of trees have the desired number of branches. After radius pruning, all bifurcations with terminal daughter branches have at least one daughter with $r\geq\tau_r$. This method is illustrated in Figure \ref{fig:Strahlerorder}(d).
  
\subsection{Directional complexity}
    
\noindent We use Topological data analysis (TDA) to examine the spatial features of  labeled spatial trees. TDA provides an extensive toolbox of data analysis methods for learning the shape of data. In this study, we use persistent homology to analyze the control and HPH spatial trees \cite{Edelsbrunner83, Edelsbrunner94, Robins99}. Specifically, we compute the 0--dimensional persistent homology of a height filtration, and use this to create a \emph{barcode} and compute a topological marker known as the directional complexity. These techniques are robust to noise \cite{Amezquita20,Sazdanovic14}, making them useful for biological applications such as the one studied here. 

In the remaind of this section we provide a brief background for TDA methods illustrated on examples from this study. Computing persistent homology involves building a sequence of topological spaces upon a data set, computing the corresponding homology for each, and tracking the changes in topological summaries that describes the shape of the data through the filtration.
    
\medskip
\noindent \textbf{Simplicial homology: } Since spatial trees are 1-dimensional examples of simplicial complexes in this section we provide a brief introduction to simplicial homology. A $k$-dimensional simplex, for $k \geq 0$, is a convex hull of $k+1$ vertices \cite{Hatcher01}. For example, 0-simplex is a point, a 1-simplex is edge, a 2-simplex is a triangular face, and a 3-simplex is a filled tetrahedron. A \emph{simplicial complex} is the formal sum of building blocks called \emph{simplices}.  Each labeled spatial tree in this study consists of a set of points/vertices which are 0-simplices connected by a set of edges which are 1-simplices. Note that all simplices come with orientation which is consistent with the fact that the  edges of spatial trees used in this stude are directed away from the root of the tree to represent the direction of blood flow.

Next we  define a linear algebra version of simplicial homology. For a given simplicial complex $S$, the \emph{chain vector space}, $C_k(S)$ is the vector space whose basis is the set of oriented $k$-simplices in $S$ \cite{Hatcher01}. The elements $c=\sum_i a_i\sigma_i\in C_k$ are called a \emph{$k$-chains} where each $a_i$ is an integer and each $\sigma_i$ is an oriented $k$-simplex from $S$. The boundary map, $\partial_k:C_k\to C_{k-1}$, maps $k$-dimensional chains to a linear combination of their $(k-1)$-dimensional boundaries. 

Let $L_k(S)$ be the kernel of $\partial_k$ and $B_k(S)$ be the image of $\partial_{k+1}$. The $k^{th}$ \emph{simplicial homology}, $H_k(S)$, is the quotient 
  \begin{equation}
      H_k(S) = L_k(S) / B_k(S) = \ker(\partial_k)/\text{im}(\partial_{k+1}).
  \end{equation}
The $k^{th}$ \emph{Betti number}, $\beta_k$, is the dimension of the k-dimensional homology $H_k$. Intuitively, $\beta_0$ counts the number of clusters (connected components) in a simplicial complex \cite{Sazdanovic14}. Property of a simplicial complex which is counted by $\beta_k$ is often referred to as a \emph{$k$-feature}.
 Betti numbers $\beta_0$, $\beta_1,$ and $\beta_2$ count the number of clusters (disjoint connected components), (unfilled) loops $S^1$, and (hollow) spheres $S^2$, respectively. 
  
\medskip
\noindent \textbf{Persistent homology:}  
Persistent homology combines ideas from topology and statistics and extends the computations to point clouds. 
The idea behind persistent homology is to compute homology of a sequence of nested simplicial complexes built from data points based on their proximity.
Given a positive parameter $\varepsilon$ and a point cloud, build a simplicial complex $\varepsilon$ by connecting all points that are less than $S_{\varepsilon_1} $ away from each other. The set of nested simplicial complexes is called a \emph{filtration} and is determined by a distance function parameter $\varepsilon$. 
Next, given an given an increasing sequence of parameter  values $\varepsilon_1,...,\varepsilon_m$ we construct a filtration of our data  $S_{\varepsilon_1}\subset S_{\varepsilon_2}\subset...\subset S_{\varepsilon_m}$.
$k$--dimensional persistent homology consists of homology of each simplex in the filtration  $H_k(S_{\varepsilon_1}),..., H_k(S_{\varepsilon_m}).$ 
The advantages of persistent homology stem from enabling the analysis of the evolution of their $k^{th}$ Betti numbers and the corresponding k-features describing the shape of data. 
If a feature appears for the first time in a certain $S_{\varepsilon_i}$, but is no longer present in $S_{\varepsilon_j}$ for some $j>i$, then the \emph{birth} of that feature is $\varepsilon_B=\varepsilon_i$ and the \emph{death} is $\varepsilon_D=\varepsilon_j$. The \emph{persistence} of a feature is $P=\varepsilon_D-\varepsilon_B$. More persistent features are ones present for a large range of $\varepsilon$ values. Note that for a large enough The last complex created in a filtration will have the topology of a point.
  
\begin{figure}[!ht]
    \centering
    \includegraphics[width=\textwidth]{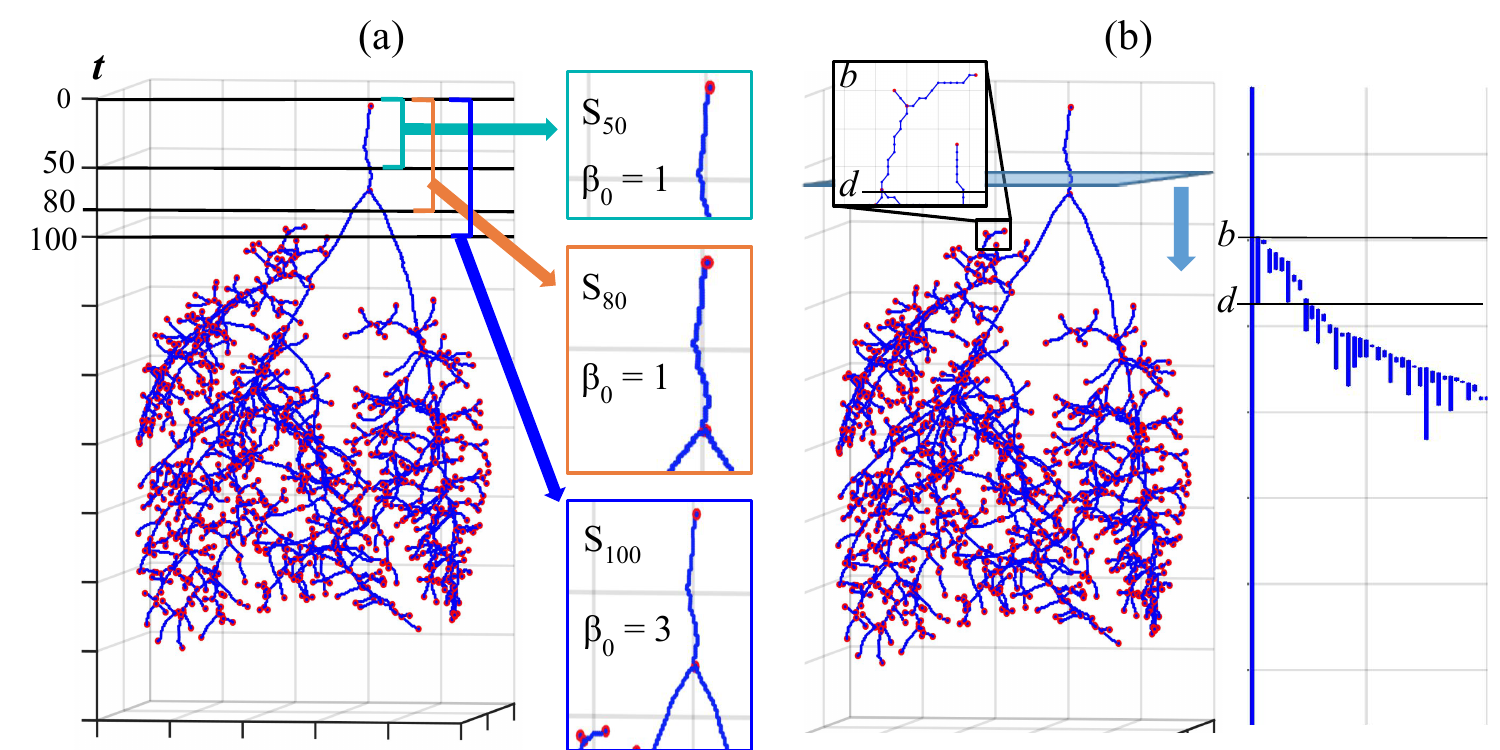}
    \caption{Directional filtration in the $-z$ direction illustrated on a labeled spatial tree. Panel (a) shows a spatial tree from a control mouse with several examples of the simplicial complexes $S_t$, where $t$ denotes the distance from the top of the tree, each labeled with their 0-dimensional Betti numbers. $S_{50}$ and $S_{80}$ each contain one connected component, so $\beta_0=1$ for both. $S_{100}$ includes 3 connected components, so $\beta_0=3$. Panel (b) illustrates the filtration process in the $-z$ direction and shows a portion of the barcode. Each time a new connected component emerges in some $S_t$, a new bar is added to the barcode. For example, the magnified branch would create a new connected component in $S_b$, where $t=b$ is called the \emph{birth} of this component. Then, in $S_d$ the branch reconnects to the main tree, no longer being a separate connected component. This means $t=d$ is the \emph{death} of this component and the bar representing this component is ended. The \emph{persistence} of this feature is the length of its bar, $d-b$. The \emph{directional complexity} (specifically \emph{anterior complexity} for $-z$ filtration) is the sum of the persistences of all these bars (equation (\ref{eq:DC})). The longest and leftmost bar in the barcode begins with the root vessel and indicates that the entire tree comprises one connected components. }
    \label{fig:complexity}
\end{figure}

\medskip
\noindent \textbf{Barcodes:} The evolution of the homology $H_k(S_{\varepsilon_i})$ can be visualized by creating a $k$-dimensional  barcode $\EuScript{B}_k$, a diagram which contains 1 bar for each new $k$-feature detected in the filtration. Each bar spans the length from the birth $\varepsilon_B$ to the death  $\varepsilon_D$ for that feature, see Figure \ref{fig:complexity}. Though the interpretations of lengths of bars in the barcode are application specific, shorter bars are often considered  to represent noise. In our case, see Figure \ref{fig:complexity}(b) for the 0-dimensional barcode of one of the spatial trees, the longest vertical bar on the far left represents the whole spatial tree, since at the end of filtration all of the branches will be connected to the root.

\begin{figure}[t!]
    \centering
    \includegraphics[width=\textwidth]{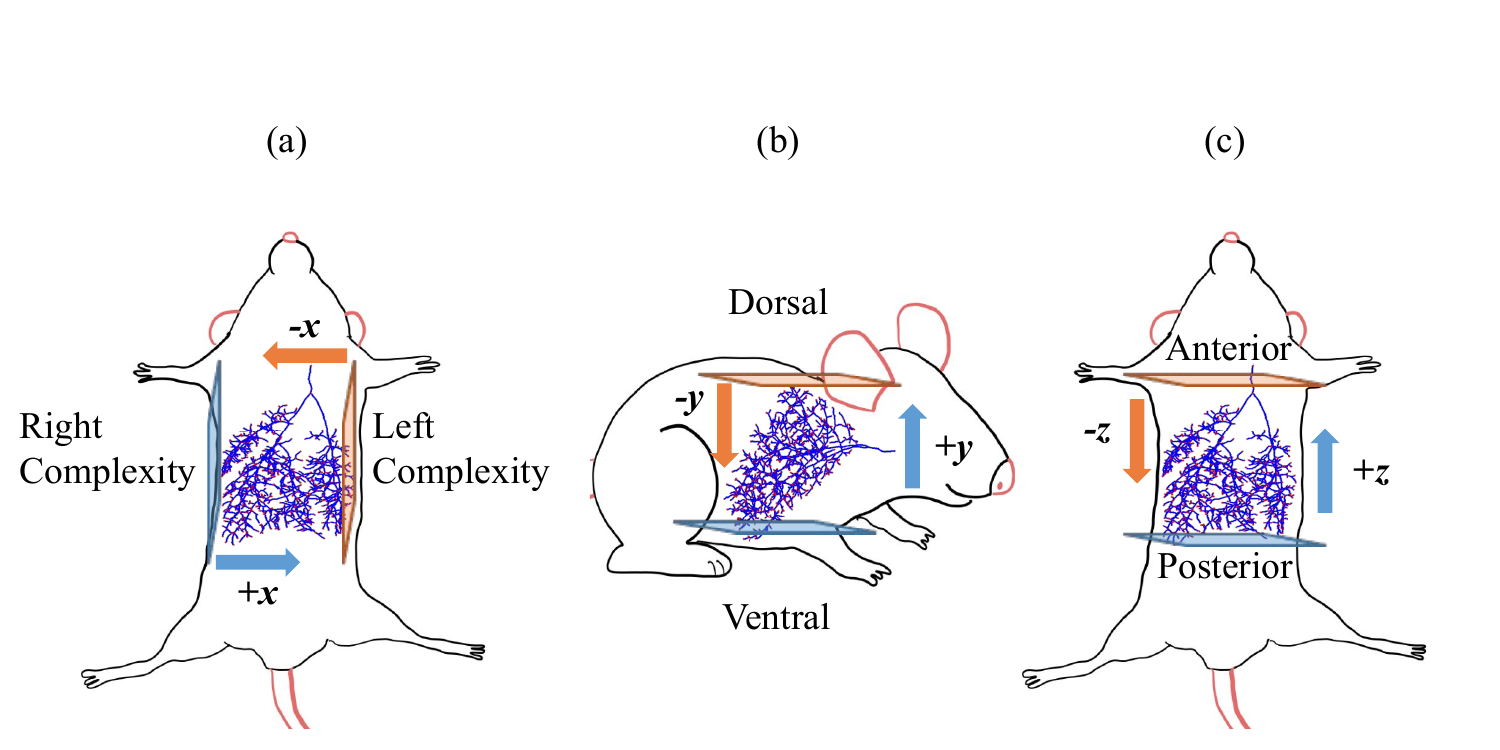}
    \caption{The 6 height filtration directions, illustrated relative to a mouse's body. The directional complexities are named after the direction of the branches they capture, which will be the reverse of the direction in which the filtration moves.}
    \label{fig:mouse}
\end{figure}

\medskip
\noindent \textbf{Height filtration:} We compute the 0-dimensional persistent homology of the spatial trees with respect to the height filtration.  Given a spatial tree $T \in \mathbb{R}^3$ and its bounding box placed so that the minimum value of the coordinate corresponding to the direction is equal to zero and consider the distance of any  vertex $V$ or a point on some edge  $E$ of $T$ to one of the sides of the box along the chosen  direction.  In general,  the height filtration requires that each $S_{\varepsilon_i}$ includes all vertices $v\in V$ and edges $e\in E$ for which the value of the distance function $\varepsilon(\cdot)$ is at most $\varepsilon_i$. The formula for  $\varepsilon(\cdot)$ depends on the direction we want to examine in our filtration. We consider six filtration directions: the positive/negative $x$-directions ($+x$/$-x$),  the positive/negative $y$-directions ($+y$/$-y$), the positive/negative $z$-directions ($+z$/$-z$) (shown in Figure \ref{fig:mouse}). For a given filtration direction the height $\pm\xi$, $\varepsilon$ of the vertex $v\in V$ with spatial coordinates $(x_v, y_v, z_v)$ is defined as
\begin{align*}
    \varepsilon(v)=\begin{cases}
			\xi_v, & \text{if direction = $+\xi$}\vspace{2mm}\\
            \xi_{\max}^T - \xi_v, & \text{if direction = $-\xi$,}\vspace{2mm}
		 \end{cases}
\end{align*} 
where $\xi_{\max}^T$ is the maximum $\xi$ coordinate of any $v\in V$ for $\xi=x$, $y$, or $z$. For an edge $e\in E$ between two vertices $v_1$ and  $v_2$, $$\varepsilon(e)=\max\left\{\varepsilon(v_1), \varepsilon(v_2)\right\}.$$

\noindent For example, a height filtration in the $+z$ direction indicates that the filtration moves along the $z$-axis in the positive direction, and each $S_{\varepsilon_i}$ includes all vertices with $z_v\leq\varepsilon_i$ and all edges with $\max\left\{z_{v_1}, z_{v_2}\right\}\leq\varepsilon_i$. Conversely, a direction of $-z$ indicates that the filtration moves along the $z$-axis in the negative direction, and each $S_{\varepsilon_i}$ includes all vertices with $z_{\max}^T - z_v\leq\varepsilon_i$ and all edges with $\max\left\{z_{\max}^T - z_{v_1}, z_{\max}^T - z_{v_2}\right\}\leq\varepsilon_i$. An illustration of the $-z$ height filtration is shown in Figure \ref{fig:complexity}.
    
\medskip 
\noindent \textbf{Directional complexity:} Given a spatial tree  $T$ and a filtration direction $\pm\xi$, the 0--dimensional barcode is computed by counting  the number of disjoint connected components in the nested set of subgraphs at each $\varepsilon_i$, Figure \ref{fig:complexity}. The directional complexity (DC) of $T$ in each direction is the total persistence of all $0$-features in the filtration, computed by summing the lengths of the bars in the barcode,
     \begin{align}
         \displaystyle \text{DC} = \sum_{\substack{k \text{ bars in} \\ \text{barcode}}} (d_k-b_k). \label{eq:DC}
     \end{align}
Directional complexities are named relative to murine anatomy and captures the length and occurrence of branches in the named direction (Figure \ref{fig:mouse}). For example, the $-z$ filtration yields anterior complexity which refers to branches pointing toward the head of the mouse (Figure \ref{fig:complexity}).

\section{Results} 

\noindent 
Data consists of mouse lung arteries imaged at the four different contrast pressures: 6.3, 7.4, 13.0, and 17.4 mmHg. Results are reported in Tables \ref{tab:graphstats} and \ref{tab:dc} as the mean and standard deviation unless otherwise stated. Table \ref{tab:graphstats} contains spatial tree statistics of the original and pruned trees including the number of branches, number of leaves, tree depth, and Strahler order. Table \ref{tab:dc} shows the trees' directional complexities generated from the 0--dimensional persistent homology with respect to the height filtrations.

\subsection{Spatial tree statistics} 

\begin{table}
    \caption{Spatial tree statistics from micro-CT images at 4 different contrast pressures (6.3, 7.4, 13.0, and 17.2 mmHg). Abbreviations used are ``C" for control trees, ``HPH" for unpruned HPH trees, ``HPH$^\text{R}$" for radius-pruned HPH trees, ``HPH$^\text{M}$" for maximal Strahler-order pruned HPH trees, and ``C$^\text{m}$" and ``HPH$^\text{m}$" for the minimal Strahler-order pruned control and HPH trees, respectively. The Strahler order is the same for all trees in each group, except for the HPH$^\text{R}$ trees at 7.4 \& 13.0 mmHg, where one tree had $SO_{MPA}=7$, and the other two had $SO_{MPA}=6$. All other values are reported as mean $\pm$ standard deviation. *At 17.2 mmHg, the HPH$^\text{M}$ and C$^\text{m}$ are the same as their uncropped versions.\\}
    \label{tab:graphstats}
    \centering
    {\small \begin{tabular}{c|c|c|c|c|c} \hline
    Pressure & Type & \# Branches & \# Leaves & Tree depth & Strahler order   \\ \hline
       & C  & $1018\pm151$ & $513\pm77$ & $25.3\pm1.5$ & $6$\\
    & HPH  & $2196\pm778$ & $1107\pm395$ &  $32.3\pm4.0$ & $7$ \\
    6.3 &HPH$^\text{R}$ & $1022\pm53$ & $516\pm29$ & $30.7\pm3.5$ & $6$ \\ 
    &HPH$^\text{M}$ & $1519\pm530$ & $766\pm270$ & $31.3\pm4.0$ & $6$ \\
    &C$^\text{m}$ & $829\pm262$ & $418\pm132$  & $24.7\pm1.5$ & $6$ \\
    &HPH$^\text{m}$ & $729\pm178$ & $368\pm91$  & $28.7\pm3.5$ & $6$ \\\hline
    &   C  & $1122\pm193$ & $565\pm98$ & $28.0\pm5.2$ & $6$ \\
    &HPH  & $2362\pm769$ & $1193\pm390$ &  $31.7\pm2.9$ & $7$ \\
    7.4 &HPH$^\text{R}$ & $1122\pm82$ & $567\pm42$ & $30.0\pm1.7$ & [6 6 7] \\ 
    &HPH$^\text{M}$ & $1337\pm346$ & $676\pm175$ & $30.0\pm2.6$ & $6$ \\
    &C$^\text{m}$ & $837\pm289$ & $422\pm146$ &  $27.0\pm5.5$ & $6$ \\
    &HPH$^\text{m}$ & $728\pm106$ & $367\pm53$ & $27.7\pm2.1$ & $6$ \\ \hline
    &   C  & $1693\pm372$ & $854\pm191$ &  $30.3\pm1.5$ & $6$ \\
    &HPH  & $2730\pm972$ & $1379\pm495$ &  $33.0\pm3.6$ & $7$ \\
     13.0&HPH$^\text{R}$ & $1689\pm206$ & $855\pm108$ & $32.0\pm2.6$ & [6 6 7] \\ 
    &HPH$^\text{M}$ & $1394\pm306$ &  $706\pm156$ & $31.0\pm3.0$ & $6$ \\
    &C$^\text{m}$ & $961\pm171$ & $485\pm84$ & $28.7\pm2.5$ & $6$ \\
    &HPH$^\text{m}$ & $725\pm84$ & $367\pm43$ & $28.3\pm2.5$ & $6$  \\ \hline
    &   C  & $2573\pm517$ & $1301\pm263$ & $32.0\pm4.4$ & $7$ \\
    &HPH  & $3239\pm1103$ & $1639\pm564$ & $35.0\pm4.0$ & $7$ \\
    17.2 &HPH$^\text{R}$ & $2592\pm563$ & $1312\pm289$ & $34.7\pm3.5$ & $7$ \\ 
    &HPH$^\text{M}$ * & $3239\pm1103$ & $1639\pm564$ & $35.0\pm4.0$ & $7$ \\
    &C$^\text{m}$ * & $2573\pm517$ & $1301\pm263$ & $32.0\pm4.4$ & $7$ \\
    &HPH$^\text{m}$ & $1955\pm304$ & $989\pm156$ & $33.7\pm3.5$ & $7$  \\ \hline
    \end{tabular}}
\end{table}
\noindent The results in Table \ref{tab:graphstats} indicate that the number of branches, number of leaves, and tree depth are all lower in the original control trees than in the original HPH trees. All original HPH trees have Strahler order $SO_{MPA}=7$, while control trees had $SO_{MPA}=6$ at all pressures except 17.2 mmHg, for which they had $SO_{MPA}=7$. This can be explained by the fact that HPH causes vessels to dilate \cite{Chambers20} (indicated by a higher length-to-radius ratio), so more vessels are detected in the micro-CT images, leading to more branches in the spatial trees. Due to higher pulmonary arterial compliance in control patients, at higher pressures the number of branches in the control trees becomes closer to that of the hypertensive trees.

\begin{figure}[ht]
    \centering
    \includegraphics[width=0.9\textwidth]{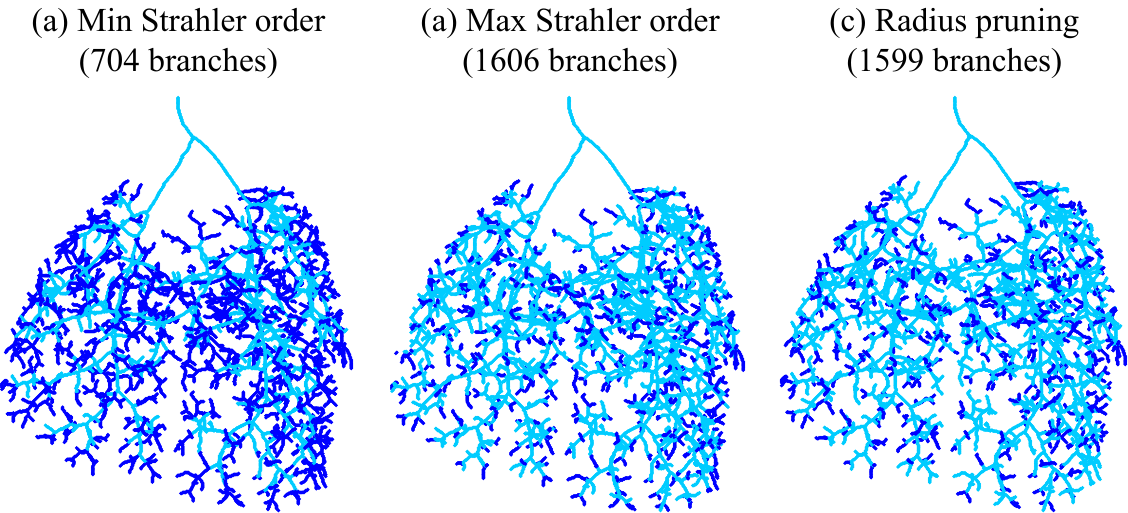}
    \caption{Pruned trees, shown in light blue, overlaid on their original trees in dark blue. The pruned trees are obtained via the 3 pruning algorithms defined in Section \ref{sec:prune}: (a) minimum Strahler order pruning, (b) maximum Strahler order pruning, and (c) radius pruning.}
    \label{fig:pruning}
\end{figure}

Normalized trees, see Figure \ref{fig:pruning}, are obtained by maximum Strahler order pruning to the HPH trees at all pressures except 17.2 mmHg, since the control and HPH trees already had the same Strahler order ($SO_{MPA}=6$) at that pressure. At the two lowest pressures (6.3 and 7.4 mmHg) these pruned HPH trees still have more branches, leaves, and a higher tree depth than the control trees. At 13.0 mmHg, the pruned HPH trees had fewer branches and leaves than the control trees, which could be attributed to increased vessel stiffness in HPH \cite{Chambers20}. Because the control vessels are more compliant, they expand more at higher pressure than HPH vessels, allowing the imaging contrast to perfuse more vessels. At the highest pressure (17.2 mmHg), maximum Strahler order pruning did not affect the HPH trees, since they already had the same Strahler order as the controls.

Minimum Strahler order pruning is applied to both control and HPH trees, generating trees of Strahler order $SO_{MPA}=6$ for pressures 6.3, 7.4, and 13.0 mmHg, and Strahler order $SO_{MPA}=7$ for the highest pressure, 17.2 mmHg. At the two lowest pressures (6.3 and 7.4 mmHg), the pruned control trees have more branches and leaves, but lower tree depth than the pruned HPH trees. At 13.0 mmHg, the pruned control trees have more branches, leaves, and higher depth than the pruned HPH trees. 
Since all control trees already have Strahler order equal to seven at the highest pressure, 17.2 mmHg, minimum Strahler order pruning does not change the original control trees, 
and the pruned HPH trees have more branches, leaves, and higher depth than control. 

Radius pruning is applied to HPH trees only, bringing the number of branches and leaves close to those of control trees. On average, the radius-pruned HPH trees have slightly more branches than the control trees at 6.3 and 17.2 mmHg, fewer branches at 13.0 mmHg, and the same number of branches at 7.4 mmHg. The radius-pruned HPH  trees have a few more leaves than the control trees at all pressures. The depth of radius-pruned HPH trees is higher than that of control trees at all pressures. The Strahler order of all radius-pruned HPH  trees are approximately the same as the control trees. 

\begin{table}
    \centering
        \caption{Directional complexities from micro-CT images at 4 different contrast pressures (6.3, 7.4, 13.0, and 17.2 mmHg). Abbreviations used are ``C" for control trees, ``HPH" for unpruned HPH trees, ``HPH$^\text{R}$" for radius-pruned HPH trees, ``HPH$^\text{M}$" for maximal Strahler-order pruned HPH trees, and ``C$^\text{m}$" and ``HPH$^\text{m}$" for the minimal Strahler-order pruned control and HPH trees, respectively. All values are reported as mean $\pm$ standard deviation. *At 17.2 mmHg, the HPH$^\text{M}$ and C$^\text{m}$ are the same as their uncropped versions.\\}
    \label{tab:dc}
  {\footnotesize   \begin{tabular}{c|c|c|c|c|c|c} \hline 
     & \multicolumn{6}{c}{Directional complexity}\\ \hline
  Type & Right & Left & Ventral & Dorsal & Posterior & Anterior \\ \hline
  \multicolumn{7}{c}{Pressure 6.3 mmHg}\\ \hline
      C & $3145\pm235$ & $2679\pm530$ & $2546\pm456$ & $3329\pm494$ & $4502\pm551$ & $1648\pm423$ \\
      HPH & $5028\pm1641$ & $5222\pm1600$ & $4460\pm1341$ & $5843\pm1495$ & $6716\pm1339$ & $3659\pm1369$ \\
      HPH$^\text{R}$ & $2582\pm338$ & $2717\pm153$ & $2160\pm85$ & $3198\pm25$ & $3788\pm155$ & $1678\pm54$  \\
      HPH$^\text{M}$ & $3745\pm1249$ & $3840\pm1160$ & $3189\pm925$ & $4383\pm1126$ & $5043\pm980$ & $2601\pm950$ \\
      C$^\text{m}$ & $2627\pm645$ & $2249\pm820$ & $2120\pm553$ & $2801\pm949$ & $3836\pm830$ & $1363\pm645$ \\
      HPH$^\text{m}$ & $1941\pm595$ & $1968\pm394$ &  $1530\pm246$ & $2431\pm402$ & $2930\pm274$ & $1195\pm293$ \\ \hline
  \multicolumn{7}{c}{Pressure 7.4 mmHg} \\ \hline
      C  & $3357\pm299$ & $2887\pm421$ & $2784\pm559$ & $3534\pm264$ & $4718\pm638$ & $1810\pm288$ \\
      HPH  & $5311\pm1634$ & $5549\pm1418$ & $4765\pm1143$ & $6135\pm1209$ & $7070\pm1262$ & $3904\pm1221$ \\
      HPH$^\text{R}$ & $2814\pm386$ & $2908\pm61$ & $2339\pm118$ & $3435\pm158$ & $4074\pm159$ & $1852\pm110$ \\
      HPH$^\text{M}$ & $3313\pm731$ & $3427\pm815$ & $2804\pm636$ & $3984\pm700$ & $4593\pm767$ & $2302\pm613$ \\
      C$^\text{m}$ & $2527\pm807$ & $2244\pm843$ & $2084\pm753$ & $2724\pm975$ & $3742\pm1021$ & $1353\pm654$ \\
      HPH$^\text{m}$ & $1917\pm256$ & $1991\pm414$ & $1544\pm313$ & $2432\pm204$ & $2980\pm318$ & $1197\pm207$ \\ \hline
  \multicolumn{7}{c}{Pressure 13.0 mmHg} \\ \hline
      C & $4254\pm706$ & $3962\pm872$ & $3771\pm500$ & $4563\pm1075$ & $5821\pm628$ & $2728\pm906$ \\ 
      HPH  & $5912\pm1859$ & $6203\pm2241$ & $5368\pm1397$ & $6797\pm1481$ & $7700\pm1600$ & $4445\pm1426$ \\
      HPH$^\text{R}$ & $4130\pm314$ & $4035\pm530$ & $3454\pm218$ & $3771\pm218$ & $4752\pm156$ & $2828\pm233$ \\
      HPH$^\text{M}$ & $3397\pm441$ & $3547\pm840$ & $2926\pm733$ & $4059\pm711$ & $4643\pm793$ & $2361\pm529$ \\
      C$^\text{m}$ & $2702\pm450$ & $2367\pm349$ & $2270\pm508$ & $2879\pm364$ & $3812\pm619$ & $1493\pm223$ \\
      HPH$^\text{m}$ & $1894\pm256$ & $1973\pm259$ & $1489\pm190$ & $2400\pm50$ & $2886\pm173$ & $1183\pm75$\\ \hline
  \multicolumn{7}{c}{Pressure 17.2 mmHg} \\ \hline
      C  & $5677\pm796$ & $5462\pm737$ & $5197\pm531$ & $6110\pm1135$ & $7400\pm487$ & $4040\pm955$ \\
     HPH  & $6619\pm2148$ & $7002\pm2064$ & $6134\pm1655$ & $7647\pm1854$ & $8455\pm1891$ & $5191\pm1659$  \\
    HPH$^\text{R}$ & $5623\pm1284$ & $5797\pm1091$ & $5039\pm761$ & $6476\pm944$ & $4208\pm811$ & $7256\pm922$ \\
     HPH$^\text{M}$ * & $6619\pm2148$ & $7002\pm2064$ & $6134\pm1655$ & $7647\pm1854$ & $8455\pm1891$ & $5191\pm1659$  \\
     C$^\text{m}$ * & $5677\pm796$ & $5462\pm737$ & $5197\pm531$ & $6110\pm1135$ & $7400\pm487$ & $4040\pm955$  \\
     HPH$^\text{m}$ & $4432\pm749$ & $4578\pm704$ & $3913\pm449$ & $5232\pm608$ & $5854\pm505$ & $3213\pm530$  \\ \hline
    \end{tabular}}
\end{table}

\subsection{Directional complexity}

\noindent All directional complexities are higher in the original HPH trees than in the original control trees. In the maximum Strahler order pruned HPH trees, all directional complexities are higher than controls at 6.3 mmHg. At 7.4 mmHg, right and posterior complexities are higher in control trees than in the radius-pruned HPH trees; all other directional complexities are higher in the radius-pruned HPH trees at this pressure.  At 13.0 mmHg, all directional complexities are lower in the pruned HPH trees than controls. Again, this may be due to higher vessel stiffness in HPH mice \cite{Chambers20}. The control vessels are more compliant, so they dilate more at higher pressure than HPH vessels, resulting in more branches in the spatial trees. When more branches point in all directions, directional complexities are higher.

For the minimum Strahler order pruned trees, all directional complexities are higher in pruned control trees than in pruned HPH trees for pressures 6.3, 7.4, and 13.0 mmHg. At 17.2 mmHg, right, ventral, and posterior complexities are lower in pruned HPH trees than controls, while left, dorsal, and anterior complexities are higher in pruned HPH trees. 

In original, maximum, and minimum Strahler pruned trees, the group (control or HPH) with the more branches at each pressure generally has higher directional complexities, with few exceptions mentioned above. The radius-pruned HPH trees, by design, have similar branch counts to the controls. At all pressures, the radius-pruned HPH trees have lower right, ventral, and posterior complexities and higher left and anterior complexities than controls. This indicates that more (and/or longer) branches point in the left and anterior directions in HPH, possibly due to the arterial network adapting to distribute blood more effectively these regions. Since mice are quadrupeds, dorsal complexity captures branches pointing in the opposite direction of gravity. Dorsal complexity is lower in radius-pruned HPH trees than controls at all pressures except 17.2 mmHg. This result agrees  with the finding in Belchi et al. \cite{Belchi18} that upwards (toward the head, also opposite of gravity) complexity in human bronchial networks decreases with COPD, a disease commonly associated with HPH. 

\section{Discussion}

\noindent This study characterizes control and HPH spatial trees, quantifying the number of branches and leaves, tree depth, Strahlerorder, and directional complexities. 

\medskip
\noindent\textbf{Tree Depth: } We found that the tree depth is higher in HPH trees compared to control, a phenomenon which is preserved after pruning the trees. While pulmonary vascular remodeling observed in HPH has been characterized by increased vessel stiffness and dilation of large arteries \cite{Chambers20}, one effect that is not typically observed is angiogenesis, the development of new blood vessels. Studies show that HPH animals have fewer arterioles (arteries with radius $\leq50$ $\mu m$) than control \cite{Vanderpool11}. Since the contrast in the micro-CT images travels through dilated vessels, more vessels are captured in the segmented arteries of HPH mice compared to the control. Consequently, the increased number of branches in the HPH trees is an artifact of the imaging process and does not translate to the HPH mice having more pulmonary arteries. Additionally, the minimum radius of the vessels captured during segmentation is approximately $39$ $\mu$m, with the vast majority of vessels having a radius greater than $50$ $\mu$m (97.3\% of vessels). Since arterioles typically have radii of $25-50$ $\mu$m, they are not captured in the images, explaining why we do not observe microvascular rarefaction described in \cite{Vanderpool11}.

\medskip
\noindent\textbf{Tree Pruning: } The radius pruned HPH trees have higher left and anterior complexity compared to the controls. This means that in HPH, more vessels point toward the head and the left. This may indicate that the network is adapting to effectively transport blood to these regions. Overall, the directional complexities correlate with number of branches. However, the HPH spatial trees have more branches due to increased vessel radius, which allows the contrast to perfuse more arteries. By design, the radius-pruning algorithm makes the number of branches in HPH trees close to that of controls. Therefore, the directional complexity results from these trees are more significant, as they will be less influenced by branch count. Moreover, the control trees have higher right, ventral, dorsal, and posterior complexities than the radius-pruned HPH trees, with the exception of dorsal complexity at the highest pressure (17.2 mmHg). This exception reflects our previous finding that higher pressure trees are more difficult to distinguish.

\medskip
\noindent\textbf{CT Image Pressure: } At higher pressures, it is more difficult to distinguish control and hypertensive trees. However, at higher contrast pressures (13.0 and 17.2 mmHg) the control and HPH trees become more similar. This can be explained by the fact that in control animals, the vessels are more compliant and expand more when perfused at higher pressures. The latter is also observed with the minimal Strahler order pruning at the second highest pressure (13.0 mmHg), where the control trees' depth exceeds that of the HPH trees.

\medskip
\noindent\textbf{Directional complexity: } Our use of 0--dimensional persistent homology via the height filtration is inspired by Belchi et al. \cite{Belchi18}, who found that humans with mild to moderate COPD had lower upwards complexity of the bronchial network compared to healthy subjects. In this context, ``upwards" complexity refers to bronchi pointing toward the head. Since HPH is the type of PH most commonly associated with COPD \cite{Galie15} and the pulmonary arteries branch in a manner similar to the bronchi, we implemented the same filtration as \cite{Belchi18} on the murine spatial trees. Belchi et al. \cite{Belchi18} found that upwards complexity of the bronchi is lower in disease; this is contrary to our findings that anterior complexity is higher in HPH. However, mice are quadrupeds and their lungs are rotated compared to humans. Thus, dorsal complexity might be a more apt comparison to human upwards complexity because it is the direction which counters gravity. We found that the dorsal complexity is lower in the radius pruned HPH trees compared to control trees at the three lowest pressures. 

Finally, unlike in the study of  Belchi et al. \cite{Belchi18} we modeled our methods on, we found that the choice of filtration direction matters in mice vascular networks while it does not influence results for the human airways. In particular,  upwards complexity was the only one reported in \cite{Belchi18} while this study detects  differences in some of the directional complexities in the pruned trees, which could be reflective of the differences between human and mouse lung structures. 
Note also that study had more data but the trees had fewer generations and lower depth while we had fewer scans but generated larger trees with up to about 35 generations. 

\medskip
\noindent\textbf{Future work and limitations } Future work include expanding TDA methods to analyze human lungs, and overcoming the  major limitation faced in this study - lack of data. Lack of data is a common problem in physiological studies where it can be difficult to conduct experiments with large number of animals, or obtaining data due to privacy concerns of patients. The  lack of data can be remedied using generative machine learning to construct surrogate networks representative of the actual data, an approach we plan to pursue in future studies. 
This approach was used in the  persistent homology study by Bendich et al.  \cite{Bendich16} analyzing brain arterial networks. Using machine learning they generated 98 trees from repeated iterations of a tube-tracking algorithm. As a first step towards  fitting the variables in generative models we will use simulated networks obtained by attaching self-similar structured trees to principal branches of  labeled spatial trees \cite{Chambers20} and incorporating angles extracted from our spatial trees to capture the  the three-dimensional properties of arterial branching.

Another challenge in analyzing human lungs is segmenting clinical CT images because they have lower resolution and in vivo images also contain airways and veins, making it difficult to identify the arteries. For a clinical CT image it is possible to identify ~2-300 vessels compared to the ~2000 vessels visibile in the micro-CT images analyzed here. The mouse images analyzed here were ideal because the micro-CT images were obtained at high resolution and the pulmonary arteries were excised, i.e., the arterial networks were the only anatomical objects in the images.

\section{Conclusion}

\noindent This study serves as a proof-of-concept for the use of TDA to detect differences in diseased vascular networks. In this work we  lay the foundation and explore the possibilities of identifying numerical features of vascular networks  based on topological data analysis of their shape  relevant for detecting HPH. Data consists of spatial trees extracted from  micro-CT images of the pulmonary arterial networks in control and HPH mice. There is large discrepancy between branch counts between the two groups, which we hypothesize is an imaging consequence rather than evidence of angiogenesis, as microvascular rarefaction has been observed in mice with HPH \cite{Vanderpool11}. Therefore, we used used pruning algorithms on trees in order to obtain biologically relevant results. Trees were pruned based on the vessel radius to account for the larger thickness of vessels in HPH mice and according to the Strahler order which reflects the asymmetry of the constructed trees. In addition to using the standard graph statistics we use numerical summary derived from persistent homology called directional complexity \cite{Belchi18} to characterize and compare pulmonary arterial networks. For example, we conclude that tree depth is larger in the HPH trees, directional complexity correlates with the branch count, and the left and dorsal complexities are lower in HPH. Finally, we noticed that results at low perfusion pressures reveal more differences, likely because at higher pressure increased elasticity in healthy animals cause vessels to dilate more making them appear similar to HPH animals.

Next steps include overcoming the  major limitation faced in this study - lack of data using generative machine learning methods and preliminary simulations from \cite{Chambers20}. Most exciting research direction is expanding TDA methods to analyze human lungs, the vascular network together with  the  airways as their structures and shapes are intertwined and dependent. develop jointly and affect i and overcoming the  major limitation faced in this study - lack of data.

\section{Acknowledgements}

\noindent This study was initiated during a summer research experience for undergraduates at North Carolina State University, funded by the NSA: H98230-20-1-0259, H98230-21-1-0014, H98230-22-1-0006. Olufsen was supported in part by NSF-DMS 2051010, 1615820. RS was partially supported by the NSF-DMS $1854705$. The authors also thank Naomi Chesler, University of California-Irvine, for providing micro-CT images for this study.

\end{document}